\documentclass{rnaastex}

\begin{document}

\title{Pre-airburst Orbital Evolution of Earth's Impactor 2018 LA: An Update}

\correspondingauthor{Carlos~de~la~Fuente~Marcos}
\email{nbplanet@ucm.es}

\author[0000-0003-3894-8609]{Carlos~de~la~Fuente~Marcos}
\affiliation{Universidad Complutense de Madrid \\
             Ciudad Universitaria, E-28040 Madrid, Spain}

\author[0000-0002-5319-5716]{Ra\'ul~de~la~Fuente~Marcos}
\affiliation{AEGORA Research Group \\
             Facultad de Ciencias Matem\'aticas \\
             Universidad Complutense de Madrid \\
             Ciudad Universitaria, E-28040 Madrid, Spain}


\section{} 

Meteoroid 2018~LA has become only the third natural object ever to be discovered prior to causing a meteor 
airburst\footnote{\href{https://minorplanetcenter.net/mpec/K18/K18L04.html}{MPEC 2018-L04: 2018 LA}} and just the second one to have its
meteorites recovered.\footnote{\href{https://seti.org/fragment-impacting-asteroid-recovered-botswana-0}{Fragments found in Botswana's 
Central Kalahari Game Reserve}} Asteroid 2008~TC$_{3}$ was the first one \citep{2009Natur.458..485J,2009A&A...507.1015B}, 2014~AA followed 
\citep{2016Ap&SS.361..358D,2016Icar..274..327F}, although in this case infrasound recordings were the only physical evidence left behind by 
the disintegration. Meteorites were recovered from 2008~TC$_{3}$ in northern Sudan \citep{2009Natur.458..485J}. The pre-impact orbital 
evolution of 2008~TC$_{3}$ was studied by \citet{2009Natur.458..485J}, \citet{2012MNRAS.424..508G}, \citet{2012P&SS...73...30O}, and 
\citet{2017Icar..294..218F}; that of 2014~AA has been discussed by \citet{2016Ap&SS.361..358D} and \citet{2016Icar..274..327F}. The case of 
2018~LA has been explored by \citet{2018RNAAS...2b..57D}, although a preliminary orbital solution was used. Here, we present an updated 
analysis based on the most recent data.

The latest orbit determination of 2018~LA (epoch JD~2458200.5, 23-March-2018, solution date 18-July-2018) is based on 28 observations (the
previous one was based on 14) and has semimajor axis, $a=1.3728\pm0.0008$~au, eccentricity, $e=0.4299\pm0.0004$, inclination, 
$i=4\fdg281\pm0\fdg002$, longitude of the ascending node, $\Omega=71\fdg8801\pm0\fdg0006$, and argument of perihelion, 
$\omega=256\fdg052\pm0\fdg011$.\footnote{\href{http://ssd.jpl.nasa.gov/sbdb.cgi}{JPL's Small-Body Database}} We have used this improved 
solution to search for additional bodies moving in paths comparable to that of 2018~LA as we did in \citet{2018RNAAS...2b..57D}, using the 
$D$-criteria, which are metrics to study orbit similarity (see e.g. \citealt{2016Ap&SS.361..358D}). We confirm that the known asteroids with 
the closest short-term evolution to that of 2018~LA are (454100) 2013~BO$_{73}$ ($D_{\rm R}=0.024$), 2016~LR ($D_{\rm R}=0.0055$), and 
2018~BA$_{5}$ ($D_{\rm R}=0.010$). As discussed in \citet{2018RNAAS...2b..57D}, the present-day orbit of 2018~BA$_{5}$ ($a=1.3670\pm0.0009$~au, 
$e=0.4310\pm0.0005$, $i=4\fdg539\pm0\fdg007$, $\Omega=338\fdg47\pm0\fdg02$, $\omega=42\fdg62\pm0\fdg02$) matches well ---in terms of size, 
$a$, and shape, $e$--- that of 2018~LA prior to entering our atmosphere. Although there are several other asteroids with orbits that 
resemble those of the four discussed here, none of them has values of the $D$-criteria under 0.005.

\begin{figure}[!ht]
\begin{center}
\includegraphics[scale=0.4,angle=0]{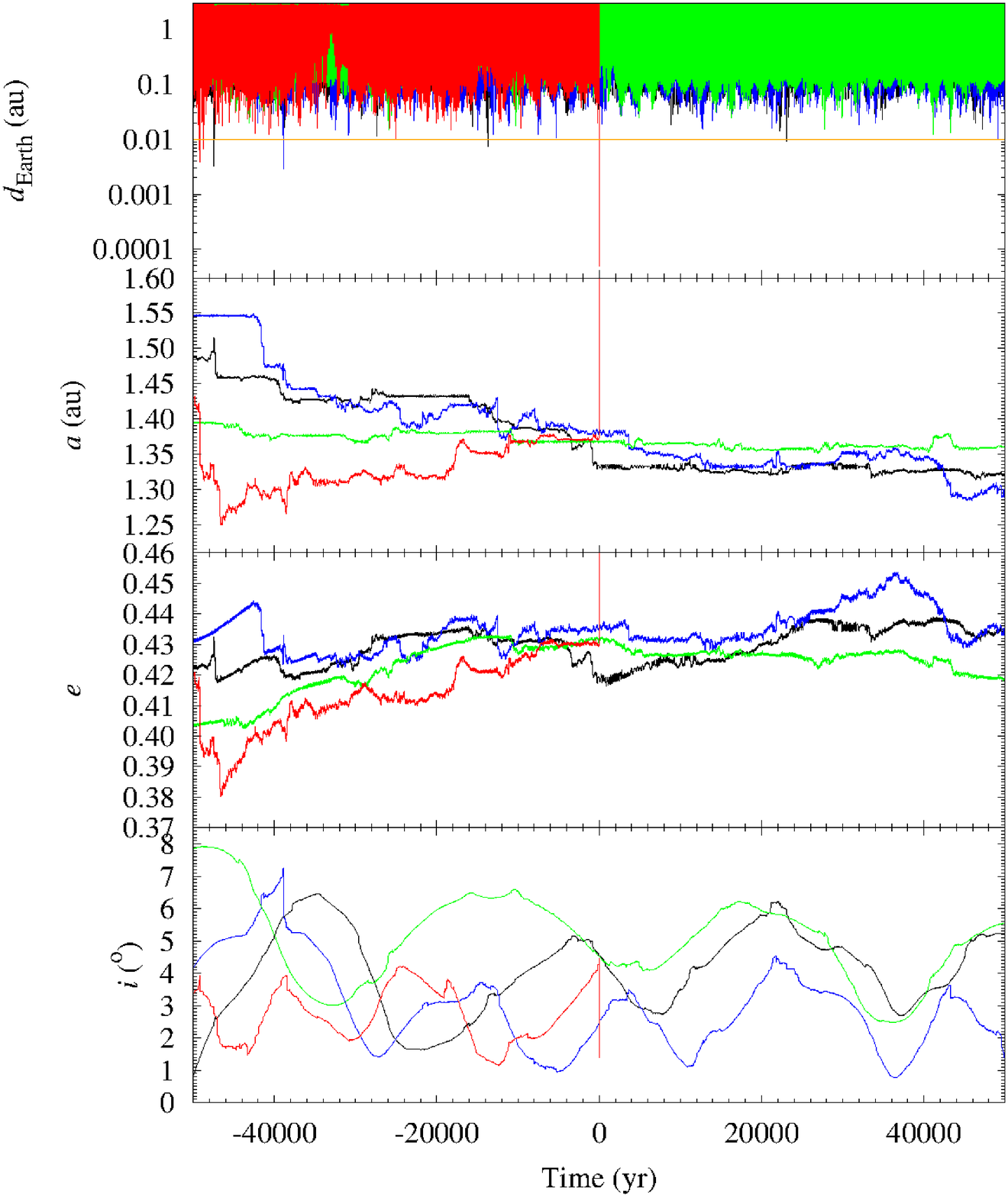}
\includegraphics[scale=0.4,angle=0]{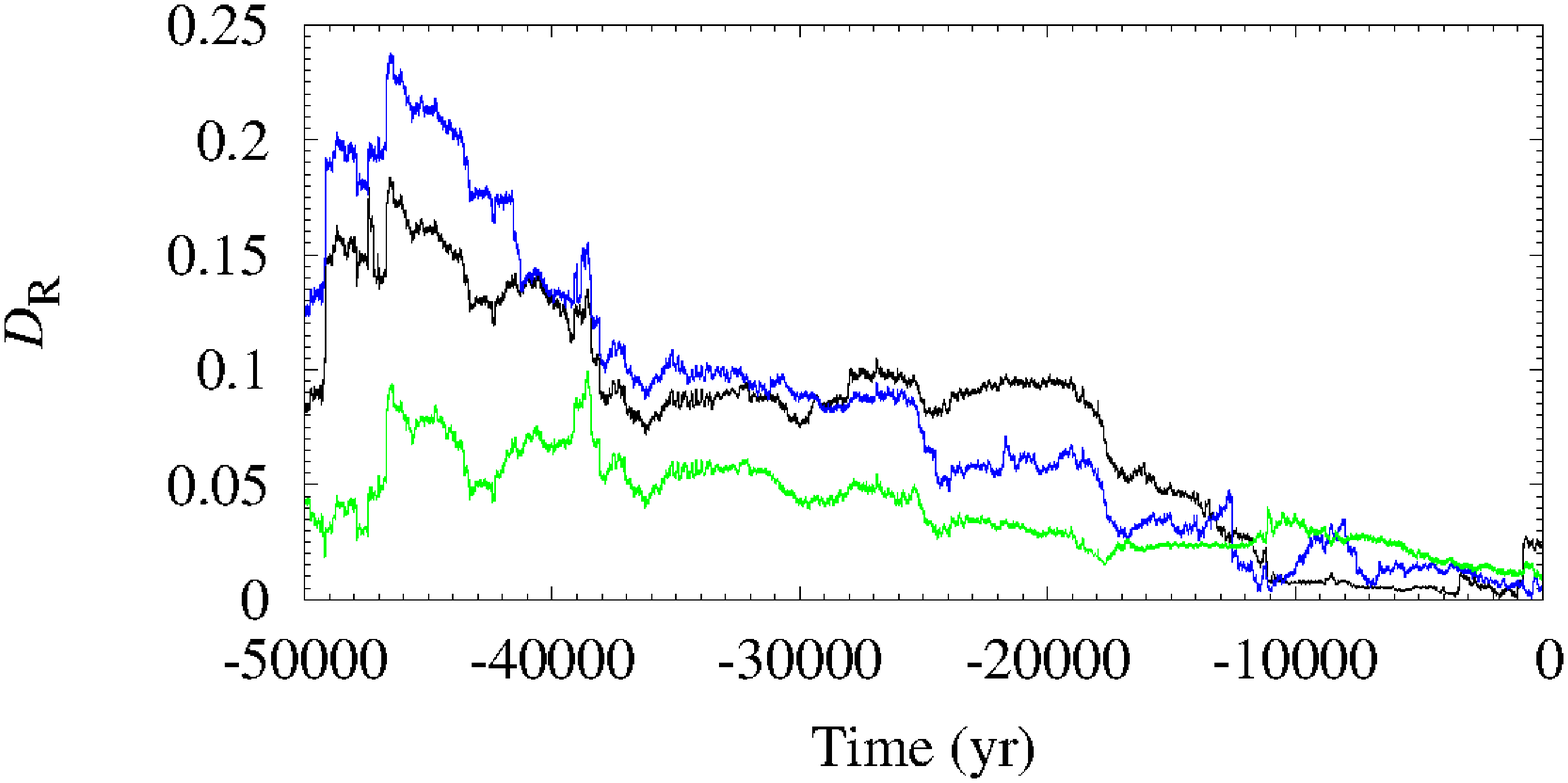}
\caption{Top panel, perigee distance (Hill radius of the Earth, 0.0098~au, in orange), second to top panel, semimajor axis, third to top 
         panel, eccentricity, second to bottom panel, inclination, and bottom panel, $D_{\rm R}$ for the nominal orbits (zero instant of 
         time, epoch JD~2458200.5~TDB, 23-March-2018) of 454100 (black), 2016~LR (blue), 2018~BA$_{5}$ (green), and 2018~LA (red).  
\label{fig:1}}
\end{center}
\end{figure}

Figure~\ref{fig:1} is an improved version of figure~1 in \citet{2018RNAAS...2b..57D} that includes the $D_{\rm R}$ (bottom panel) and shows 
the pre-impact orbital evolution of 2018~LA and the three asteroids mentioned before. The comparison of both figures shows that the behavior 
of 2018~LA and in general of objects moving in similar orbits is very sensitive to initial conditions. Such a chaotic response makes it 
difficult to reconstruct precisely their dynamical evolution beyond a few hundred years. This property is also shared by previous impactors 
\citep{2009Natur.458..485J,2012MNRAS.424..508G,2012P&SS...73...30O,2016Ap&SS.361..358D,2016Icar..274..327F,2017Icar..294..218F} and it is a 
side effect of experiencing flybys with the Earth--Moon system and other inner planets. The evolution of the value of $D_{\rm R}$ for 454100 
and 2018~LA during the last 10\,000 years or so is intriguing enough (Figure~\ref{fig:1}, bottom panel), but if a recent fragmentation 
episode is not a valid scenario able to explain these similarities, being part of a dynamical grouping \citep{2016MNRAS.456.2946D} could be 
a feasible alternative. 

In this Note, we have provided an updated discussion of the pre-impact orbital evolution of 2018~LA, which is the parent body of the 
fireball observed over South Africa and Botswana on 2018 June 2 and also of the meteorites eventually found in Botswana's soil. We have 
further confirmed the existence of a group of asteroids that might be related to each other. The improved data suggest that 2018~LA could be 
a recent fragment spawned by a larger object, like 454100 (550~m, Table 4 in \citealt{2016AJ....152...63N}). Spectroscopic observations of 
454100 during its next flyby with our planet (brightest at an apparent visual magnitude of 18.4 on 2018 mid-November) may confirm or deny a 
putative similar chemical composition to that of the recovered meteorites of 2018~LA.

\acknowledgments

We thank S.~J. Aarseth for providing the code used in this research and A.~I. G\'omez de Castro for providing access to computing facilities. 
This work was partially supported by the Spanish MINECO under grant ESP2015-68908-R. In preparation of this Note, we made use of the NASA 
Astrophysics Data System and the MPC data server.


\begin{thebibliography}{}

\bibitem[Borovi{\v c}ka \& Charv{\'a}t(2009)]{2009A&A...507.1015B} Borovi{\v c}ka, J., \& Charv{\'a}t, Z.\ 2009, \aap, 507, 1015

\bibitem[de la Fuente Marcos \& de la Fuente Marcos(2016)]{2016MNRAS.456.2946D} de la Fuente Marcos, C., \& de la Fuente Marcos, R.\ 2016, \mnras, 456, 2946

\bibitem[de la Fuente Marcos et al.(2016)]{2016Ap&SS.361..358D} de la Fuente Marcos, C., de la Fuente Marcos, R., \& Mialle, P.\ 2016, \apss, 361, 358

\bibitem[de la Fuente Marcos \& de la Fuente Marcos(2018)]{2018RNAAS...2b..57D} de la Fuente Marcos, C., \& de la Fuente Marcos, R.\ 2018, Research Notes of the American Astronomical Society, 2, 57

\bibitem[Farnocchia et al.(2016)]{2016Icar..274..327F} Farnocchia, D., Chesley, S.~R., Brown, P.~G., \& Chodas, P.~W.\ 2016, \icarus, 274, 327

\bibitem[Farnocchia et al.(2017)]{2017Icar..294..218F} Farnocchia, D., Jenniskens, P., Robertson, D.~K., et al.\ 2017, \icarus, 294, 218

\bibitem[Gayon-Markt et al.(2012)]{2012MNRAS.424..508G} Gayon-Markt, J., Delbo, M., Morbidelli, A., \& Marchi, S.\ 2012, \mnras, 424, 508

\bibitem[Jenniskens et al.(2009)]{2009Natur.458..485J} Jenniskens, P., Shaddad, M.~H., Numan, D., et al.\ 2009, \nat, 458, 485

\bibitem[Nugent et al.(2016)]{2016AJ....152...63N} Nugent, C.~R., Mainzer, A., Bauer, J., et al.\ 2016, \aj, 152, 63

\bibitem[Oszkiewicz et al.(2012)]{2012P&SS...73...30O} Oszkiewicz, D., Muinonen, K., Virtanen, J., Granvik, M., \& Bowell, E.\ 2012, \planss, 73, 30

\end{thebibliography}
\end{document}